\documentclass[aps,prl,preprint,superscriptaddress,showpacs,showkeys]{revtex4}
\usepackage{graphicx}% Include figure files
\usepackage{dcolumn}% Align table columns on decimal point
\usepackage{bm}% bold math
\usepackage[normalem]{ulem}

\begin{document}

\title{Angular Distributions for \boldmath ${}^{3,4}_{\,\,\Lambda}\mathrm {H} $\unboldmath~Bound States in the~\boldmath$^{3,4}\mathrm {He}(e,e'K^+)$\unboldmath reaction}

\author{F.\ Dohrmann}
\email{F.Dohrmann@fz-rossendorf.de}
\affiliation{Argonne National Laboratory, Argonne, Illinois 60439}
\affiliation{Forschungszentrum Rossendorf, 01314 Dresden, Germany}
\author{A.~Ahmidouch}
\affiliation{Hampton University, Hampton, Virginia 23668}
\affiliation{Kent State University, Kent, Ohio 44242}
\author{C.S.~Armstrong}
\affiliation{Thomas Jefferson National Accelerator Facility, Newport News, Virginia 23606}
\affiliation{College of William and Mary, Williamsburg, Virginia 23187}
\author{J.~Arrington}
\affiliation{Argonne National Laboratory, Argonne, Illinois 60439}
\author{R.~Asaturyan}
\affiliation{Yerevan Physics Institute, Yerevan, Armenia}
\author{S.~Avery}
\affiliation{Hampton University, Hampton, Virginia 23668}
\author{K.~Bailey}
\affiliation{Argonne National Laboratory, Argonne, Illinois 60439}
\author{H.~Bitao}
\affiliation{Hampton University, Hampton, Virginia 23668}
\author{H.~Breuer}
\affiliation{University of Maryland, College Park, Maryland 20742}
\author{D.S.~Brown}
\affiliation{University of Maryland, College Park, Maryland 20742}
\author{R.~Carlini}
\affiliation{Thomas Jefferson National Accelerator Facility, Newport News, Virginia 23606}
\author{J.~Cha}
\affiliation{Hampton University, Hampton, Virginia 23668}
\author{N.~Chant}
\affiliation{University of Maryland, College Park, Maryland 20742}
\author{E.~Christy}
\affiliation{Hampton University, Hampton, Virginia 23668}
\author{A.~Cochran}
\affiliation{Hampton University, Hampton, Virginia 23668}
\author{L.~Cole}
\affiliation{Hampton University, Hampton, Virginia 23668}
\author{J.~Crowder}
\affiliation{Juniata College, Huntingdon, Pennsylvania 16652}
\author{S.~Danagoulian}
\affiliation{North Carolina A\&T State University, Greensboro, North Carolina 27411}
\affiliation{Thomas Jefferson National Accelerator Facility, Newport News, Virginia 23606}
\author{M.~Elaasar}
\affiliation{Southern University at New Orleans, New Orleans, Louisiana 70126}
\author{R.~Ent}
\affiliation{Thomas Jefferson National Accelerator Facility, Newport News, Virginia 23606}
\author{H.~Fenker}
\affiliation{Thomas Jefferson National Accelerator Facility, Newport News, Virginia 23606}
\author{Y.~Fujii}
\affiliation{Tohoku University, Sendai, 980-8577 Japan}
\author{L.~Gan}
\affiliation{Hampton University, Hampton, Virginia 23668}
\author{K.~Garrow}
\affiliation{Thomas Jefferson National Accelerator Facility, Newport News, Virginia 23606}
\author{D.F.~Geesaman}
\affiliation{Argonne National Laboratory, Argonne, Illinois 60439}
\author{P.~Gueye}
\affiliation{Hampton University, Hampton, Virginia 23668}
\author{K.~Hafidi}
\affiliation{Argonne National Laboratory, Argonne, Illinois 60439}
\author{W.~Hinton}
\affiliation{Hampton University, Hampton, Virginia 23668}
\author{H.~Juengst}
\affiliation{University of Minnesota, Minneapolis, Minnesota 55455}
\author{C.~Keppel}
\affiliation{Hampton University, Hampton, Virginia 23668}
\author{Y.~Liang}
\affiliation{Hampton University, Hampton, Virginia 23668}
\author{J.H.~Liu}
\affiliation{University of Minnesota, Minneapolis, Minnesota 55455}
\author{A.~Lung}
\affiliation{Thomas Jefferson National Accelerator Facility, Newport News, Virginia 23606}
\author{D.~Mack}
\affiliation{Thomas Jefferson National Accelerator Facility, Newport News, Virginia 23606}
\author{P.~Markowitz}
\affiliation{Florida International University, Miami, Florida 33199}
\affiliation{Thomas Jefferson National Accelerator Facility, Newport News, Virginia 23606}
\author{J.~Mitchell}
\affiliation{Thomas Jefferson National Accelerator Facility, Newport News, Virginia 23606}
\author{T.~Miyoshi}
\affiliation{Tohoku University, Sendai, 980-8577 Japan}
\author{H.~Mkrtchyan}
\affiliation{Yerevan Physics Institute, Yerevan, Armenia}
\author{S.K.~Mtingwa}
\affiliation{North Carolina A\&T State University, Greensboro, North Carolina 27411}
\author{B.~Mueller}
\affiliation{Argonne National Laboratory, Argonne, Illinois 60439}
\author{G.~Niculescu}
\affiliation{Hampton University, Hampton, Virginia 23668}
\affiliation{Ohio University, Athens, Ohio 45701}
\author{I.~Niculescu}
\affiliation{Hampton University, Hampton, Virginia 23668}
\affiliation{The George Washington University, Washington DC 20052}
\author{D.~Potterveld}
\affiliation{Argonne National Laboratory, Argonne, Illinois 60439}
\author{B.A.~Raue}
\affiliation{Florida International University, Miami, Florida 33199}
\affiliation{Thomas Jefferson National Accelerator Facility, Newport News, Virginia 23606}
\author{P.E.~Reimer}
\affiliation{Argonne National Laboratory, Argonne, Illinois 60439}
\author{J.\ Reinhold}
\affiliation{Florida International University, Miami, Florida 33199}
\affiliation{Thomas Jefferson National Accelerator Facility, Newport News, Virginia 23606}
\author{J.~Roche}
\affiliation{College of William and Mary, Williamsburg, Virginia 23187}
\author{M.~Sarsour}
\affiliation{University of Houston, Houston, Texas 77204}
\author{Y.~Sato}
\affiliation{Tohoku University, Sendai, 980-8577 Japan}
\author{R.E.~Segel}
\affiliation{Northwestern University, Evanston, Illinois 60201}
\author{A.~Semenov}
\affiliation{Kent State University, Kent, Ohio 44242}
\author{S.~Stepanyan}
\affiliation{Yerevan Physics Institute, Yerevan, Armenia}
\author{V.~Tadevosian}
\affiliation{Yerevan Physics Institute, Yerevan, Armenia}
\author{S.~Tajima}
\affiliation{Duke University and Triangle Universities Nuclear Laboratory, Durham, North Carolina 27708}
\author{L.~Tang}
\affiliation{Hampton University, Hampton, Virginia 23668}
\author{A.~Uzzle}
\affiliation{Hampton University, Hampton, Virginia 23668}
\author{S.~Wood}
\affiliation{Thomas Jefferson National Accelerator Facility, Newport News, Virginia 23606}
\author{H.~Yamaguchi}
\affiliation{Tohoku University, Sendai, 980-8577 Japan}
\author{C.~Yan}
\affiliation{Kent State University, Kent, Ohio 44242}
\author{L.~Yuan}
\affiliation{Hampton University, Hampton, Virginia 23668}
\author{B.~Zeidman}
\affiliation{Argonne National Laboratory, Argonne, Illinois 60439}
\author{M.~Zeier}
\affiliation{University of Virginia, Charlottesville, Virginia 22901}
\author{B.~Zihlmann}
\affiliation{University of Virginia, Charlottesville, Virginia 22901}

\begin{abstract}
  The \ensuremath{^3_{\Lambda}\mathrm{H}\;} and
  \ensuremath{^4_{\Lambda}\mathrm{H}\;} hypernuclear bound states have been
  observed for the first time in kaon electroproduction on
  \ensuremath{^{3,4}\mathrm{He}} targets. The production cross sections
    have been determined at $Q^2= 0.35 \; \mathrm{GeV}^2$ and $W=
    1.91 \; \mathrm{GeV}$.  For either hypernucleus the nuclear form factor
    is determined by comparing the angular distribution of the
    \ensuremath{^{3,4}\mathrm{He}(e,e'K^+)^{3,4}_{\Lambda}\mathrm{H}} processes to the elementary cross section \ensuremath{^1\mathrm{H}(e,e´K^+)\Lambda} on the free proton, measured during the same experiment.
\end{abstract}

\pacs{21.45.+v, 21.80.+a, 25.30.Rw, 27.10.+h}

\maketitle

\begin{figure*}
\includegraphics[width=8cm]{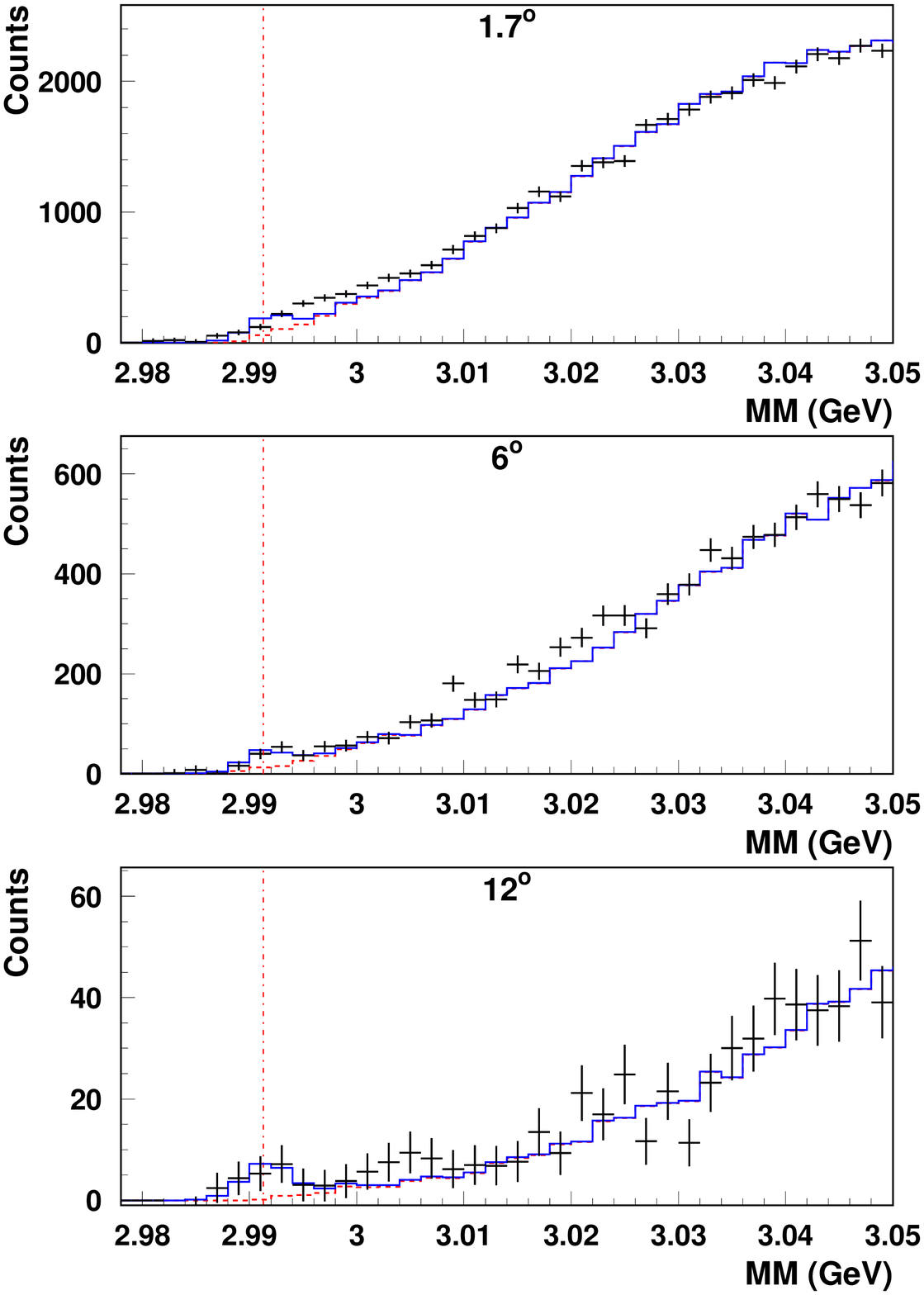}
\includegraphics[width=8cm]{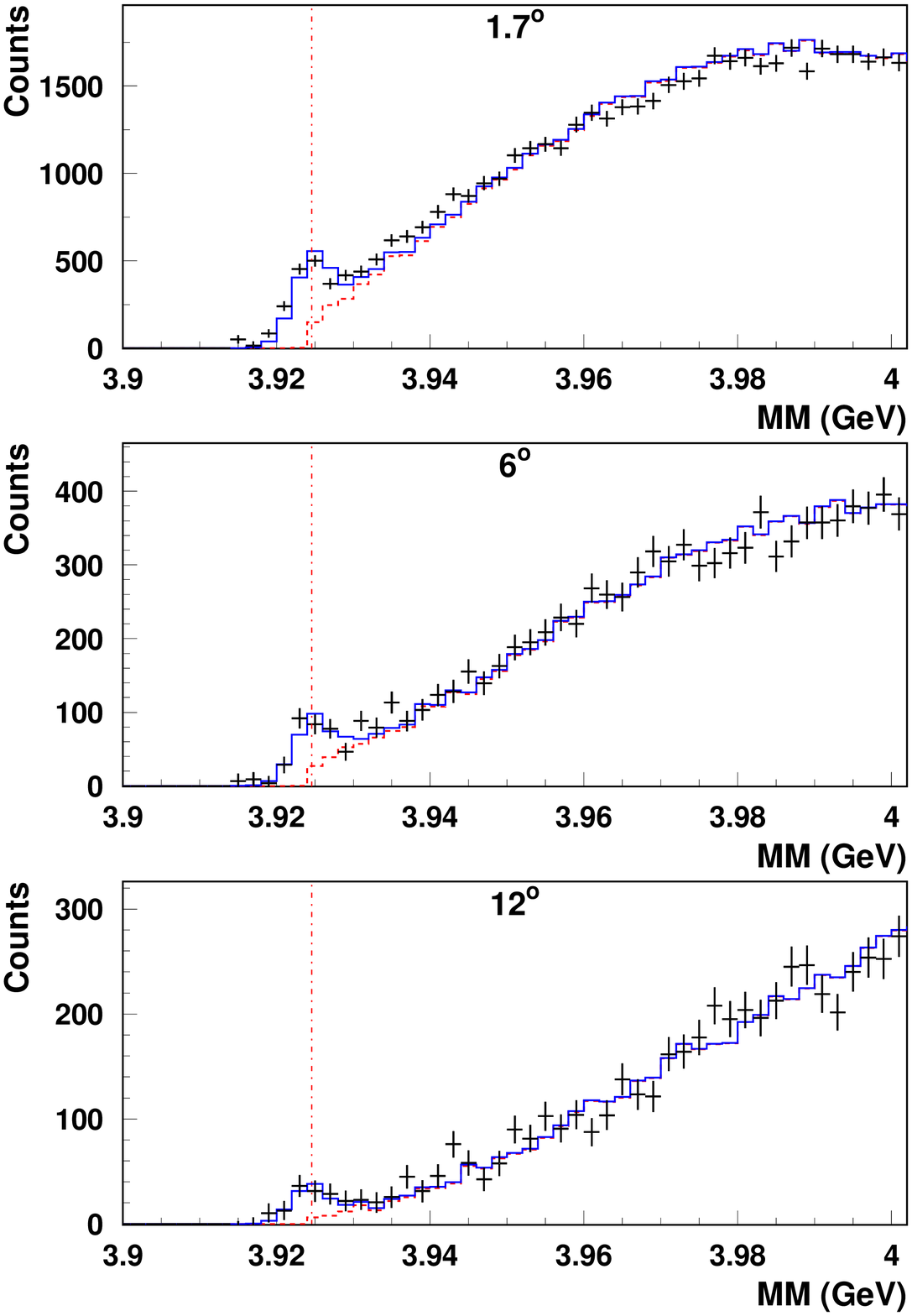}%
\caption{\label{he40612} Reconstructed missing mass spectra for $^3 \mathrm{He}$ (left) and $^4 \mathrm{He}$ (right) targets in the region
of quasifree $\Lambda$ production for different kinematic settings. Data points are shown with statistical error
  bars. Simulations of the quasifree (q.f.) reactions \ensuremath{{}^{3,4}\mathrm{He}(e,e'K^+)} are shown by dashed lines. Solid lines represent the sum of simulations of the q.f.\ background and the bound state reactions,\ensuremath{{}^{3,4}\mathrm{He}(e,e'K^+)^{3,4}_{\Lambda}\mathrm{H}}.   The thresholds for q.f.\ production, $\Lambda+^2 \mathrm{H}$, $\Lambda+^3\mathrm{H}$, respectively, are denoted by vertical lines.}
\end{figure*}

This paper presents the first reported results in the measurement of
angular distributions of electroproduced hypernuclear bound states on
${}^{3,4}\mathrm{He}$, namely the \ensuremath{^3_{\Lambda}\mathrm{H}\;} and \ensuremath{^4_{\Lambda}\mathrm{H}\;} bound states.
In a hypernucleus, one of the nucleons has been replaced by a hyperon,
i.e.\ $\Lambda$ or $\Sigma$, so that the hyperon inside the nucleus
carries strangeness in contrast to the remaining nucleons. This new
degree of freedom inside the nucleus is not blocked by the
Pauli-Principle. Inasmuch as hypernuclei provide a laboratory in which
to study the strong hyperon--nucleon interaction as well as the weak
decay of the hyperons in the nuclear medium, hyperons within a nucleus
may also be viewed as impurities probing the nuclear structure
\cite{Gibson:1995an}.

There is no known bound hyperon-nucleon system for $A=2$. The
hypertriton \ensuremath{^3_{\Lambda}\mathrm{H}\;} is the lightest hypernucleus and the only one with
$A=3$. For $A=4$, both \ensuremath{^4_{\Lambda}\mathrm{H}\;} and $^4_{\Lambda}\mathrm{He}$ are
bound. These light hypernuclei were first observed more than 50 years
ago as hyperfragments in emulsion studies \cite{Danysz1956}.  Since
these early measurements, these hypernuclei have not been studied in
reaction spectroscopy, inasmuch as \ensuremath{^3_{\Lambda}\mathrm{H}\;} and \ensuremath{^4_{\Lambda}\mathrm{H}\;} cannot be
produced from $\mathrm{He}$ targets in reactions employing only
charged meson beams and ejectiles, e.g.\ the established $(\pi^+,K^+)$
and $(K^-,\pi^-)$ reactions. The advent of high quality and high
intensity electron beams offers a novel opportunity to study these
nuclei in the $(e,e'K^+)$ reaction.

The results presented here are part of a study of kaon
electroproduction on light nuclei, E91016, conducted in Hall C of the Thomas Jefferson National
Accelerator Facility. The data were obtained at an electron
beam energy of $3.245 \; \mathrm{GeV}$ and beam currents of $20 - 25 \;
\mu \mathrm{A}$ incident upon specially developed high density cryogenic targets for
$A=1-4$. 
The Helium target lengths were approximately $4\; \mathrm{cm}$, the thicknesses being $310\; \mathrm{mg/cm^2}\; (^3\mathrm{He})$ and $546 \; \mathrm{mg/cm^2}\;(^4\mathrm{He})$, $\pm 1\%$
respectively. The backgrounds from uncorrelated $(e',K^+)$ pairs, as well as
contributions from the aluminum walls of the cryogenic targets, were
subtracted in the charge normalized yields.

\begin{figure}
\includegraphics[width=8cm]{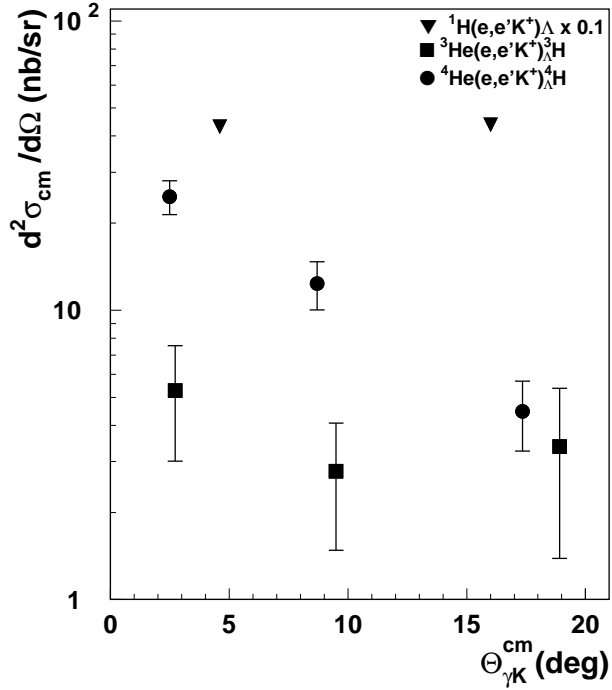}
\caption{\label{he4angular} Angular distributions of cross sections in the virtual photon-nucleus center of mass for the \ensuremath{^{3,4}\mathrm{He}(e,e'K^+)^{3,4}_{\Lambda}\mathrm{H}} and \ensuremath{^1\mathrm{H}(e,e'K^+)\Lambda} (scaled by $0.1$) processes, plotted vs.\ $\Theta^{\mathrm{cm}}_{\gamma\mathrm{K}}$. The data are given in Table \ref{values}.}
\end{figure}

The scattered electrons were detected in the High Momentum
Spectrometer (HMS, momentum acceptance $\Delta p/p \simeq \pm 10\%$,
solid angle $\sim 6.7 \; \mathrm{msr}$) 
in coincidence with the electroproduced kaons, detected in the Short
Orbit Spectrometer (SOS, momentum acceptance $\Delta p/p \simeq \pm
20\%$, solid angle $\sim 7.5\; \mathrm{msr}$).  The detector packages
of the two spectrometers are very similar\cite{Niculescu:1998zj}. Two drift chambers near the
focal plane, utilized for reconstructing the particle trajectories,
are followed by two pairs of segmented plastic scintillators that
provide the main trigger signal as well as time-of-flight information.
The time-of-flight resolution is $\sim 150\; \mathrm{ps}\,(\sigma)$.  For
electron identification, a lead-glass shower detector array is used
together with a gas threshold \v Cerenkov, in order to distinguish
between $e^-$ and $\pi^-$. For kaon identification in the SOS, a
silica aerogel detector (n=1.034) provided $K^+/\pi^+$ discrimination
while an acrylic \v Cerenkov counter (n=1.49) was used for $K^+/p$
discrimination.  Utilizing time of flight together with the \v
Cerenkov detectors, kaons are clearly separated from background pions
and protons \cite{Abbott:1998tq, Mohring:2002tr}.  Electroproduction processes exchange virtual photons, $\gamma^*$, between projectile and
target. The spectrometer angle for
electron  detection was kept fixed during the experiment, thereby
holding the virtual photon flux constant (cf.\ Ref. \cite{Zeidman:2001sh}).  The angle of the kaon arm was varied to measure angular distributions with respect to the direction of $\gamma^*$. The invariant mass of $\gamma^*$ was $Q^2 = 0.35 \; \mathrm{GeV}^2$, the total energy in the photon-nucleon system was $W= 1.91 \; \mathrm{GeV}$. The
$^{3,4}\mathrm{He}(e,e'K^+)\mathrm{X}$ process was studied for
three different angle settings between $\gamma^*$
and the ejected kaon ($K$), $\theta_{\gamma^*K^+}^{\mathrm{lab}}
\simeq 1.7^o,\; 6^o,\;12^o$,  that correspond to 
increasing the momentum transfer to the hypernucleus ($\mid t\mid \simeq
(0.12,\; 0.14,\; 0.23) \; \mathrm{GeV}^2$).
The central spectrometer momenta were $1.29\; \mathrm{GeV/c}$ for the
kaon arm and $1.58\; \mathrm{GeV/c}$ for the electron arm.

The final states, $X$, in $^{3,4}\mathrm{He}(e,e'K)X$ 
in the reconstructed missing mass spectra of the recoiling system are
shown in Fig.\ \ref{he40612}, were identified using the four-momenta $q$ of the
virtual photon, $p_K$ of the outgoing kaon, and total missing momentum $P_\mathrm{miss}$,
$M_x^2 = (q + P_\mathrm{miss} - p_K)^2$.  For $^4\mathrm{He}$, a \ensuremath{^4_{\Lambda}\mathrm{H}\;}
bound state is clearly visible for all three angles just below the
$^3\mathrm{H}-\Lambda$ threshold of $3.925\,\mathrm{MeV}$. For
$^3\mathrm{He}$, just below the $^2\mathrm{H}-\Lambda$ threshold of
$2.993\,\mathrm{MeV}$, the \ensuremath{^3_{\Lambda}\mathrm{H}\;} bound state is barely visible as a
weak shoulder for $1.7^o$, but clearly present for $6^o$ and $12^o$.

Two states of the \ensuremath{^4_{\Lambda}\mathrm{H}\;} system are known \cite{Gibson:1995an}, the
ground state with a binding energy of $(2.04 \pm 0.04)\, \mathrm{MeV}$,
$J^{\pi}=0^+$, and an excited state, bound by $(1.00 \pm 0.06)\,
\mathrm{MeV}$, $J^{\pi}=1^+$.  The experimental resolution of $\sim 4\; \mathrm{MeV}$
is, however, not sufficient to resolve the ground and excited states
of the \ensuremath{^4_{\Lambda}\mathrm{H}\;} system.  
The calibration of the missing mass spectrum
has been performed using elastic $^1\mathrm{H}(e,e'p)$ data as well as
$^1\mathrm{H}(e,e'K^+)\Lambda$ data, both obtained during the same experiment.
The precision of the calibration is estimated to be better than $1\,\mathrm{MeV}$.
Since the missing mass spectra have not been shifted with respect to the
known binding energy of the \ensuremath{^4_{\Lambda}\mathrm{H}\;} state after calibration, the observed
agreement shows the adequacy of the procedure. Inasmuch as
electroproduction has a large spin-flip probability 
in the forward direction \cite{Cotanch:1986yc}, the excited state of the
\ensuremath{^4_{\Lambda}\mathrm{H}\;} system should be favored and the data interpreted as a superposition
of the ground and excited states, where the excited state is favored even
more strongly closer to $0^o$.

\begin{figure}
\includegraphics[width=8cm]{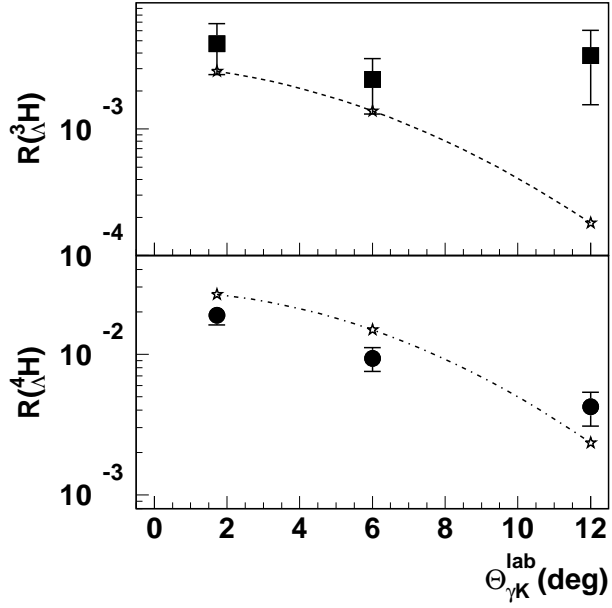}%
\caption{\label{he4formfactor} Ratios 
$R=\sigma_{\mathrm{lab}}(^{3,4}\mathrm{He})/\sigma_{\mathrm{lab}}(^1\mathrm{H})$ for $^3_{\Lambda}H$ (upper panel) and $^4_{\Lambda}H$ (lower panel). The dashed and dot-dashed curves are related to the respective nuclear form factors calculated in Ref.\ \cite{Mart1998, barz2004}.}
\end{figure}

\begin{table*}
\caption{\label{values} Differential cross sections for electroproduction of \ensuremath{^3_{\Lambda}\mathrm{H}\;} and
\ensuremath{^4_{\Lambda}\mathrm{H}\;} bound states. In the laboratory, $\sigma_{\mathrm{lab}}$ denotes the five fold differential cross section 
${d^5\sigma}/{d\Omega_e dE_e d\Omega_K }$,  $\sigma_{\mathrm{cm}}$ denotes the two fold differential cross section  
${d^2\sigma}/{d\Omega}$ 
in the virtual photon-nucleus center of mass system. The last two columns give the cross section for the $^1\mathrm{H}(e,e'K^+)\Lambda$ process, obtained during the same experiment. The combined statistical and systematic errors are given. The first row shows data for $1.7^o$ averaged over the azimuth. The $\theta_{\mathrm{cm}}$ angles corresponding to $\theta_{\mathrm{lab}}$ of $1.7^o, 6^o, 12^o$ are:\ $\;2.7^o,9.5^o,18.9^o;\;2.5^o, 8.7^o, 17.4^o;4.6^o, 16.0^o, 31.7^o$; for $^3\mathrm{He}$, $^4\mathrm{He}$, and $^1\mathrm{H}$ targets, respectively. }
\begin{ruledtabular}
\newcolumntype{d}[1]{D{.}{.}{#1}}
\begin{tabular}{cd{1.8}d{1.8}d{1.10}d{2.9}d{2.9}d{3.9}}
\multicolumn{1}{c}{\ensuremath{\theta_{\gamma^*, K^+}^{\mathrm{lab}} (^o)}} & \multicolumn{2}{c}{\ensuremath{^3_{\Lambda}\mathrm{H}}} &
\multicolumn{2}{c}{\ensuremath{^4_{\Lambda}\mathrm{H}}} & \multicolumn{2}{c}{\ensuremath{\Lambda}}\\ 
\multicolumn{1}{c}{\ }& 
\multicolumn{1}{c}{\ensuremath{\sigma_{\mathrm{lab}} \mathrm{(nb/GeV/sr^2)} }} & 
\multicolumn{1}{c}{\ensuremath{\sigma_{\mathrm{cm}} \mathrm{(nb/sr)}}} & 
\multicolumn{1}{c}{\ensuremath{\sigma_{\mathrm{lab}}}}  & 
\multicolumn{1}{c}{\ensuremath{\sigma_{\mathrm{cm}} }}  & 
\multicolumn{1}{c}{\ensuremath{\sigma_{\mathrm{lab}}}}  & 
\multicolumn{1}{c}{\ensuremath{\sigma_{\mathrm{cm}} }}  \\

 \ensuremath{<1.7>} & 
  0.045 \ensuremath{\;\pm\;}   0.008 &   5.15 \ensuremath{\;\pm\;}   0.94  & 
  0.156 \ensuremath{\;\pm\;}   0.008 &  20.83 \ensuremath{\;\pm\;}   1.13  & 
 10.59 \ensuremath{\;\pm\;}   0.21 & 465.01 \ensuremath{\;\pm\;}   9.42 \\ 
  
 1.7 & 
  0.047 \ensuremath{\;\pm\;}   0.020 &   5.27 \ensuremath{\;\pm\;}   2.26  & 
  0.185 \ensuremath{\;\pm\;}   0.025 &  24.70 \ensuremath{\;\pm\;}   3.31  & 
  9.80 \ensuremath{\;\pm\;}   0.52 & 430.43 \ensuremath{\;\pm\;}  22.75 \\ 
  
 6 & 
  0.024 \ensuremath{\;\pm\;}   0.011 &   2.77 \ensuremath{\;\pm\;}   1.30  & 
  0.093 \ensuremath{\;\pm\;}   0.018 &  12.36 \ensuremath{\;\pm\;}   2.34  & 
  9.90 \ensuremath{\;\pm\;}   0.17 & 437.31 \ensuremath{\;\pm\;}   7.61 \\ 
  
 12 & 
  0.029 \ensuremath{\;\pm\;}   0.017 &   3.38 \ensuremath{\;\pm\;}   2.00  & 
  0.032 \ensuremath{\;\pm\;}   0.009 &   4.47 \ensuremath{\;\pm\;}   1.21  & 
  7.58 \ensuremath{\;\pm\;}   0.13 & 363.78 \ensuremath{\;\pm\;}   6.10 \\ 

\end{tabular}
\end{ruledtabular}
\end{table*}

The electroproduction cross section may be written as
\[ \frac{d^5\sigma}{d\Omega_e dE_e d\Omega_K }=
\Gamma\cdot\frac{d\sigma}{d\Omega_K}\quad,\] where
$\frac{d\sigma}{d\Omega_K}$ is the virtual photon cross section and
$\Gamma$ denotes the virtual photon flux factor, viz
\[ \Gamma = \frac{\alpha}{2\pi}\frac{E_e^{'}}{E_e}\frac{1}{Q^2}\frac{1}{1-\epsilon}\frac{W^2-M^2}{2M}\quad, \]
where $M$ is taken to be the nucleon mass.  The experimental cross
sections were extracted using a Monte Carlo simulation that modeled the
optical conditions of the spectrometers, kaon decays, small angle
scattering, energy losses and radiative corrections
\cite{Ent:2001hm, Mohring:2002tr}. The $^{3,4}\mathrm{He}(e,e'K^+)^{3,4}_{\Lambda}
\mathrm{H}$ bound state production process was modeled assuming
coherent production off a stationary target nucleus.  In order to
facilitate the subtraction of the unresolved quasifree tail underneath the bound
state region, the quasifree $^{3,4}\mathrm{He}(e,e'K^+)X$ processes
off nucleons inside the target nuclei, $^{3,4} \mathrm{He}$ had
to be modeled as well \cite{Uzzle2002}.  Since no models are available for the
electroproduction on $A=3,4$ nuclei, we use an elementary
cross section model \cite{Reinhold:2001zm,Mohring:2002tr} which is
convolved by spectral functions \cite{Benhar:1994hw} for $^{3,4}
\mathrm{He}$. Our dedicated model was shown to describe our
$^{1}\mathrm{H}(e,e'K^+)$ data best over the acceptance
\cite{Reinhold:2001zm}. Final state interactions in the vicinity of the
respective quasifree thresholds were taken into account by using
an effective range approximation \cite{Gillespie1964}
which gave satisfactory results as shown in Fig.~\ref{he40612}.

The uncertainty of fitting the strength of the background to the
quasifree continuum is the dominant source of the error of the cross
section for the bound state distributions, particularly for the low
yields for the $^3_{\Lambda}\mathrm{H}$ bound states, this results in
large uncertainties. Furthermore, since the effective range
approximation is very simple and takes into account only 2--body
$\Lambda$--nucleon interactions, a model dependent error has been
estimated by fitting the shape of the quasifree tail with a simple
parabolic function. This results in larger background subtractions and
leads to cross sections $\sim 20\%$ lower than for the effective range
ansatz. Thus the background was estimated to be the mean of the two
results with an additional error derived from the differences of the
two cases. Any other sources of systematic uncertainties are on a few
per cent level.
The extracted cross sections are given in Table \ref{values}. For the angular distribution shown in
Figure~\ref{he4angular}, data were restricted to a common covered range
in azimuthal angle of $(180\pm24)^o$ (last three rows of Table
\ref{values}). 
The point to point systematic uncertainties are 
 $\sim$ 36\%, 39\%, 50\%; 11\%, 16\%, 23\%; 4.4\%, 1.5\%, 1.4\% for $^3\mathrm{He}$, $^4\mathrm{He}$, $^1\mathrm{H}$, respectively.  For the setting with near parallel kinematics, $1.7^o$,
however, the full azimuth was covered (first row of
Table~\ref{values}). For these data, the point to point systematic
uncertainties are $\sim$ 15\%, 4.3\%, 4.4\% for $^3\mathrm{He}$,
$^4\mathrm{He}$, $^1\mathrm{H}$, respectively.  For both
$^{3}_{\Lambda}\mathrm{H}$ and $^4_{\Lambda}\mathrm{H}$ the cross section at
$6^o$ is a factor of two lower than at $1.7^o$. The $12^o$ data point
for $^3_{\Lambda}\mathrm{H}$, however, strongly deviates from this behaviour,
making the $^3_{\Lambda}\mathrm{H}$ angular distribution flatter than for
$^4_{\Lambda}\mathrm{H}$.

Fig.\ \ref{he4formfactor} shows the ratio
$R=\sigma_{\mathrm{lab}}(^{3,4}\mathrm{He})/\sigma_{\mathrm{lab}}(^1\mathrm{H})$
of the laboratory cross sections of
\ensuremath{^{3,4}\mathrm{He}(e,e'K^+)^{3,4}_{\Lambda}\mathrm{H}} to
the cross section on the free proton.  $R$ is related to the nuclear
form factor $F(k)$ by $R(k) = S \cdot W_A^2 \cdot F^2(k)$
\cite{Mart1998}, where $k$ is the three momentum transfer to the
hypernucleus, $S$ is a spin factor and $W_A$ combines phase space and flux factors. We
calculate $R(k)$ for our kinematics using $W_A = 2.1 (1.68)$, $k =
2.02 (2.08), 2.19 (2.23), 2.69 (2.69)\,\mathrm{fm^{-1}} $ for
$^3_{\Lambda}\mathrm{H}$ $(^4_{\Lambda}\mathrm{H})$. For
$^3_{\Lambda}\mathrm{H}$ we take the parametrization of $F(k)$ and
$S=1/6$ of Ref.\ \cite{Mart1998}, in which Gaussian approximations for
the underlying $^2\mathrm{H}$, $^3\mathrm{He}$ wave functions are
used. For $^4_{\Lambda}\mathrm{H}$ an expression of $F(k)$ similar to
the $^3_{\Lambda}\mathrm{H}$ case is derived \cite{barz2004}, using
Gaussian approximations for the underlying $^3\mathrm{H}$,
$^4\mathrm{He}$ wave functions and the charge elastic form factors of $^{3,4}\mathrm{He}$ of Ref.\ 
\cite{Wiringa1991} for parametrization, and $S=2$ for symmetry reasons. At $1.7^o$ the calculated
reduction of the elementary cross section by the form factor is $\sim$ 250 (\ensuremath{^3_{\Lambda}\mathrm{H}}) and 100 (\ensuremath{^4_{\Lambda}\mathrm{H}}). The shape of the calculated $R(k)$ for both hypernuclei is similar
to the shape of the data for $1.7^o$ and $6^o$, while it deviates
for the $12^o$ data. The latter may indicate a breakdown
of approximating the underlying wave functions by Gaussians. 
For
$^4_{\Lambda}\mathrm{H}$ at $1.7^o$ and $6^o$, the calculated $R(k)$
is $~40-50\%$ higher than the data which suggests that the underlying
wave functions are too simplistic such that their overlap is too
large. Realistic wave functions obtained from Faddeev calculations are
expected to give rise to more precise information. Future measurements
at high three momentum transfer $k$ would be highly desirable.

The production of the bound hypernuclei
\ensuremath{^3_{\Lambda}\mathrm{H}\;} (hypertriton) and
\ensuremath{^4_{\Lambda}\mathrm{H}\;} has been achieved for the first
time in electroproduction and, for the first time in reaction
spectroscopy, angular distributions for the
$^{3,4}\mathrm{He}(e,e'K)^{3,4}_{\Lambda}\mathrm{H}$ processes have
been obtained. The angular distribution for $^3\mathrm{He}$ is flatter
than for $^4\mathrm{He}$ at large angles.
Comparing these cross sections to the cross section on
the free proton shows that the angular dependence of the cross section
is determined by the nuclear form factor for small angles but deviates
for larger angles. This should be tested by performing more precise
calculations using realistic wave functions.  These data and future
measurements using dedicated spectrometer systems with resolutions
below $1\,\mathrm{MeV}$ \cite{Fujii:2003qc} may trigger a renaissance
of the spectroscopy of the lightest hypernuclei that have not been
studied since the first emulsion experiments many years ago.
 
\begin{acknowledgments}
  We are grateful to H.W.\ Barz for calculating the $\mathrm{He}$ form factors and thank T.\ Mart for valuable communications.
  This work was supported in part by the DOE under contract No.\ W-31-109-Eng-38 (ANL),
  contract No.\ DE-AC05-84ER40150 (TJNAF), and the NSF.  The
  excellent support of the staff of the Accelerator and Physics
  Division of TJNAF is gratefully acknowledged.  F.D.\ 
  acknowledges the support by the A.v.Humboldt-Stiftung through a
  Feodor Lynen-Fellowship, and the support by ANL for hosting this research. 
\end{acknowledgments}

\end{document}